\documentstyle[12pt,aps,prd,preprint]{revtex}
\begin{document}
\title{Best Approximation to a Reversible Process in Black-Hole 
Physics and the Area Spectrum of Spherical Black Holes}
\author{Shahar Hod}
\address{The Racah Institute for Physics, The
Hebrew University, Jerusalem 91904, Israel}
\date{\today}
\maketitle

\begin{abstract}
The assimilation of a {\it quantum} ({\it finite} size) particle by a
Reissner-Nordstr\"om black hole inevitably involves an increase in the
black-hole surface area. 
It is shown that this increase can be {\it
  minimized} if one considers the capture of the {\it lightest
  charged} particle in nature.
The unavoidable area increase is attributed to two physical reasons: the
{\it Heisenberg quantum uncertainty principle} and a {\it Schwinger-type 
charge emission} (vacuum polarization).
The fundamental lower bound on the area increase 
is $4 \hbar$, which is {\it smaller}
than the value given by
Bekenstein for neutral particles. Thus, this process is a better
approximation to a reversible process in black-hole physics.
The {\it universality} of the minimal area increase 
is a further evidence in favor of a {\it uniformly} spaced 
area spectrum for spherical quantum black holes. Moreover, this
universal value is in excellent agreement with the area spacing
predicted by Mukhanov and Bekenstein and independently by Hod.
\end{abstract}

\section{introduction}\label{introduction}
Can the assimilation of a test particle by a black hole be
made {\it reversible} in the sense that all changes in the black-hole parameters
can be undone by another suitable process ?
This seemingly naive question goes deep into the basic laws of black-hole physics.
A {\it classical} theorem of Hawking's \cite{Hawking} says that black-hole
surface area cannot decrease. Hence, any physical process which
increases the horizon area is obviously (classically) irreversible.
The answer to the above question was given by Christodoulou
\cite{Chris} 
(later generalized by Christodoulou and Ruffini for the case of
charged point particles \cite{ChrisRuff})
almost three decades ago. The assimilation of a ({\it point}) particle
is reversible if it is injected at the {\it horizon} from 
a {\it turning point} of its motion. In such a case the black-hole
surface area is left {\it unchanged} and the changes in the other black-hole parameters
(mass, charge and angular-momentum) can be undone by another suitable
(reversible) process.

However, as was pointed out by Bekenstein in his seminal
work \cite{Beken1} the limit of a {\it point} particle is not a legal
one in {\it quantum} theory. As a concession to quantum theory
Bekenstein ascribes to the particle a {\it finite} proper radius $b$
while continuing to assume, in the spirit of Ehrenfest's theorem, that
the particle's center of mass follows a classical trajectory. 
Bekenstein \cite{Beken1} has shown that the assimilation of the 
finite size neutral particle inevitably causes an 
increase in the horizon area.
This increase is minimized if the particle is captured when its center
of mass is at a turning point a proper distance $b$ away
from the horizon \cite{Beken1}:

\begin{equation}\label{Eq1}
(\Delta \alpha)_{min}=2\mu b\  ,
\end{equation}
where the ``rationalized area'' $\alpha$ is related to the black-hole surface area $A$ by
$\alpha = A/4 \pi$ and $\mu$ is the rest mass of the particle.
For a point particle $b=0$ and one finds $(\Delta
\alpha)_{min}=0$. This is Christodoulou's result for a reversible
process. However, a quantum particle is subjected to quantum
uncertainty. According to Bekenstein's analysis, a 
relativistic quantum particle cannot be localized to
better than its Compton wavelength (This claim certainly is not
correct when the particle has (locally measured) energy greater than
its mass. In this paper we give a different and rigorous 
argument which leads to
a lower bound on the increase in the black-hole surface area). 
Thus, $b$ can be no smaller than
$\hbar / \mu$. From here one finds a lower bound on the increase in
the black-hole surface area due to the assimilation of a (neutral)
test particle

\begin{equation}\label{Eq2}
(\Delta \alpha)_{min}=2\hbar\  .
\end{equation}
It is easy to check that the reversible process of Christodoulou and
Ruffini and the lower bound Eq. (\ref{Eq2}) of Bekenstein are valid only for {\it
  non}-extremal black holes. 
Thus, for non-extremal black holes there is a {\it universal}
(i.e., independent of the black-hole parameters) minimum area increase
as soon as one allows quantum nuances to the problem. 
This fact is used as one of
the major arguments in favor of a {\it uniformly} spaced 
area spectrum for quantum black-holes \cite{Beken2}.

The universal lower bound Eq. (\ref{Eq2}) derived by Bekenstein is valid
only for {\it neutral} particles \cite{Beken1}. 
In this paper we analyze the assimilation of 
a quantum ({\it finite} size) {\it charged} 
particle by a Reissner-Nordstr\"om black hole and show that 
the fundamental lower-bound on the increase in the black-hole surface
area is {\it smaller} than the 
value given by Bekenstein for neutral particles.

\section{Assimilation of a charged particle by a Reissner-Nordstr\"om black-hole}\label{Sec2}

The major goal of this paper is to calculate the (inevitable) {\it
  minimal} increase in black-hole surface area caused by the
assimilation of a particle of rest mass $\mu$, charge $e$ 
and proper radius $b$. 
We are interested in the area increase ascribable 
to the {\it particle} itself, as
contrasted with any increase incidental to the {\it process} of 
bringing the particle to the black-hole horizon \cite{Beken1}. 
For example,
gravitational radiation emitted by the particle \cite{Davis} or by 
any device which
might had lowered it into the hole \cite{Beken3} will also cause an 
increase in the area. 
In addition, electromagnetic radiation emitted during the process of
bringing the charged particle to the horizon \cite{Ruffini} 
will also result in an
increase in the black-hole surface area.
In this paper, as in the seminal work of Bekenstein
\cite{Beken1}, we ignore these incidental effects and
concentrate on the inevitable increase in the black-hole surface area 
caused by the captured particle all by itself.

The external gravitational field of a spherically symmetric object of 
mass $M$ and charge $Q$ is given by the Reissner-Nordstr\"om metric

\begin{equation}\label{Eq3}
ds^2=-\left( {1-{{2M} \over r}+{{Q^2} \over {r^2}}} \right)dt^2+\left( {1-{{2M}
\over r}+{{Q^2} \over {r^2}}} \right)^{-1}dr^2+r^2d\Omega ^2\  .
\end{equation}
The black hole's (event and inner) horizons are located at

\begin{equation}\label{Eq4}
r_{ \pm}=M \pm (M^2-Q^2)^{1/2}\  ,
\end{equation}

The equation of motion of a charged
particle on the Reissner-Nordstr\"om background is a quadratic equation
for the conserved energy $E$ of the particle \cite{Carter}

\begin{equation}\label{Eq5}
r^4 E^2 -2eQr^3E+ e^2Q^2r^2-
\Delta(\mu^2 r^2 +{p_{\phi}}^2)- (\Delta p_r)^2=0\  ,
\end{equation}
where $\Delta$ is given by

\begin{equation}\label{Eq6}
\Delta=r^2-2Mr+Q^2=(r-r_{-})(r-r_{+})\  .
\end{equation}
The quantities $p_{\phi}$ and $p_r$ are the conserved 
angular momentum of the particle and its
covariant radial momentum, respectively. 
It is useful to express this last quantity in terms of the physical
component (in an orthonormal tetrad) $P 
\equiv \Delta^{-1/2} r p^r$ \cite{Beken4}.

In order to find the change in the black-hole surface area caused by
an assimilation of a point particle one should first solve
Eq. (\ref{Eq5}) for $E$ and then evaluate it at the horizon $r=r_{+}$
of the black-hole.
As was pointed in Refs. \cite{Chris} and \cite{ChrisRuff}, this increase is
minimized (actually {\it vanishes}) if the particle is captured from a
turning point.
How would the {\it non}-zero proper radius $b$ of the particle (which
is an {\it inevitable} feature of the quantum theory) change this
scenario ?
First, we note that in the spirit of the Ehrenfest's theorem 
we continue to assume that the particle's center of mass follows a
classical path.
Second, as was pointed out by Bekenstein, regardless of the
manner in which the particle arrives at the horizon, it must acquire
its parameters ($E$ and $p_{\phi}$) while every part of it is still
outside the horizon, i.e., while it is not yet part of the black hole
\cite{Beken1}. Thus, the motion of the particle's center of mass at
the moment of capture should be described by Eq. (\ref{Eq5}).
Third, in order to generalize the results given in 
Refs. \cite{Chris} and \cite{ChrisRuff} to the case of a {\it
  finite} size particle one should evaluate $E$ at $r=r_{+}+ \delta (b)$,
where $\delta(b)$ is determined by \cite{Beken1}

\begin{equation}\label{Eq7}
\int_{r_{+}}^{r_{+}+ \delta (b)} (g_{rr})^{1/2} dr = b\  .
\end{equation}
In other words, $r=r_{+}+ \delta (b)$ is a point a proper distance $b$
outside the horizon.
Integrating Eq. (\ref{Eq7}) one finds

\begin{equation}\label{Eq8}
\delta (b)=(r_{+}-r_{-}) sinh^2 \left( {b \over {2r_{+}}} \right)[1+O(b/r_{+})]\  .
\end{equation}
Since we consider the case $b \ll r_{+}$ we may replace this
expression by

\begin{equation}\label{Eq9}
\delta (b)=(r_{+}-r_{-}) {b^2 \over {4{r_{+}}^2}}\  .
\end{equation}

The conserved energy $E$ of a particle having a physical radial
momentum $P$ at
$r=r_{+}+ \xi$ (where $\xi \ll r_{+}$) is given by Eq. (\ref{Eq5})

\begin{eqnarray}\label{Eq10}
E&=&{eQ \over r_{+}}+ {{\sqrt{(\mu^2+P^2) r_{+}^2 +p_{\phi}^2} 
(r_{+}-r_{-})^{1/2}} \over
  {r_{+}^2}} \xi^{1/2}\left\{1+
O\left[{\xi /(r_{+}-r_{-})}\right] \right\} \nonumber \\
&& -{eQ \over r_{+}^2} \xi \left\{1+
O\left[{\xi /(r_{+}-r_{-})}\right] \right\}\  .
\end{eqnarray}
This expression (for $P=0$) is actually 
the effective potential (gravitational
plus electromagnetic plus centrifugal) for given 
values of $\mu, e$ and $p_{\phi}$.
It is clear that it can be {\it minimized} by taking $p_{\phi}=0$ (which
also minimize the increase in the black-hole surface area. This is
also the case for neutral particles \cite{Beken1}). However, $P^2$
cannot be said to vanish because of Heisenberg quantum 
uncertainty principle. 
For $eQ > 0$ the effective potential has a {\it maximum} located at

\begin{equation}\label{Eq11}
\xi^{*}={{(r_{+}-r_{-}) (\mu^2+P^2) r_{+}^2} \over 
{4e^2Q^2}}\  .
\end{equation}

The assimilation of the particle results in a change $dM=E$ in the
black-hole mass and a change $dQ=e$ in the black-hole charge.
Using the first-law of black-hole thermodynamics

\begin{equation}\label{Eq12}
dM=\Theta d\alpha + \Phi dQ\  ,
\end{equation}
where $\Theta={1 \over 4}(r_{+}-r_{-})/\alpha$ and $\Phi=Qr_{+}/
\alpha$, one finds

\begin{equation}\label{Eq13}
d\alpha_{min}(s,\mu,e,b)={4(\mu^2 +P^2)^{1/2} r_{+}
\over {(r_{+}-r_{-})^{1/2}}}
\delta(b)^{1/2} -
{4eQ \over {r_{+}-r_{-}}} \delta(b)\  ,
\end{equation}
which is the {\it minimal} area increase for given values of the
black-hole parameters $r_{+}$ and $Q$ ($s$ stands for these two
parameters) and the particle parameters $\mu, e, b$ and $P$.

In order to be captured by the black-hole the particle has to be over
the potential barrier. There are two distinct cases that should be
treated separately: for particles satisfying the relation $\delta(b) \leq
\xi^{*}$ the area increase is {\it minimized} if $bP$ is
minimized. However, the limit $bP \to 0$ is not a legal one in
the quantum theory. According to Bekenstein's analysis \cite{Beken1} the
particle cannot be localized to better than its 
Compton wavelength $\hbar / \mu$ (We will discuss the validity of this
assumption below. 
In this paper we shall use a more rigorous argument to provide
a lower bound on the product $bP$).
On the other hand, particles
satisfying the inequality $\delta(b) > \xi^{*}$ cannot be captured
from a turning point of their motion. In order to overcome the
potential barrier and be captured by the
black hole they must have (at least) an energy $E(\xi^{*})$.

Let us consider the first case $\delta(b) \leq \xi^{*}$.
Substituting Eq. (\ref{Eq9}) into Eq. (\ref{Eq13}) 
one finds

\begin{equation}\label{Eq14}
d\alpha_{min}(s,\mu,e,b)=2(\mu^2 +P^2)^{1/2} b-{{eQb^2} 
\over {r_{+}^2}}\  .
\end{equation}
According to Bekenstein's original analysis \cite{Beken1} one may
minimize this expression by minimizing the value of $b$ 
(and setting $P^2=0$ at the turning point). However, the 
claim used in
\cite{Beken1} that a particle
cannot be localized to within less than its Compton wavelength
certainly is not correct when the particle has (locally measured) energy
greater than its mass. The locally measured energy (by a static
observer) of a particle near the horizon of a black hole can be
arbitrarily large. In addition, 
at the turning point the physical radial momentum 
$P^2$ cannot be said to vanish, but must be replaced by its 
uncertainty $(\delta P)^2$ \cite{Beken4}. 
In other words, according to Heisenberg quantum uncertainty principle 
the particle's center of mass cannot be placed at the
horizon with accuracy better than the radial position uncertainty
$\hbar/ (2\delta P)$.

Using the restriction $\delta(b) \leq \xi^{*}$ one finds

\begin{equation}\label{Eq15}
|e| \leq{{(\mu^2+P^2)^{1/2} r_{+}^2} \over {|Q|b}}\  .
\end{equation}
Thus, the minimal area increase is given by

\begin{equation}\label{Eq16}
d\alpha_{min}(s,\mu)=[\mu^2 +(\delta P)^2]^{1/2} b\  ,
\end{equation}
which, according to Heisenberg quantum uncertainty principle, yields
(for $\delta P \gg \mu$)

\begin{equation}\label{Eq17}
d\alpha_{min}(s,\mu)=\hbar /2\  .
\end{equation}

Next, we consider the assimilation of particles which satisfy the relation 
$\delta(b) > \xi^{*}$. These particles cannot be captured
from a turning point of their motion. In order to be captured by the
black hole they must have a minimal energy of

\begin{equation}\label{Eq18}
E_{min}=E(\xi^*)={eQ \over r_{+}} +{{(\mu^2+P^2)(r_{+}-r_{-})} \over
{4eQ}}\  .
\end{equation}
Using the first-law of black-hole thermodynamics Eq. (\ref{Eq12}) one
finds that the increase in the black-hole surface area is given by

\begin{equation}\label{Eq19}
d\alpha_{min}(s,\mu,e)={{(\mu^2+P^2) r_{+}^2} \over {eQ}}\  .
\end{equation}

What physics prevents us from using particles which 
make expression (\ref{Eq19}) as small as we wish ? Or, in other words,
what physics prevents us from recovering Christodoulou's 
reversible process $\Delta A=0$ ?
The answer is {\it Schwinger-type charge emission} 
(vacuum polarization) \cite{Schwin}. We must remember 
that the the black hole may
{\it discharge} itself through a Schwinger-type emission. The critical
electric field $\Xi_c$ for pair-production of particles with rest mass $\mu$
and charge $e$ is given by \cite{Schwin,ParTio,DamRuf}

\begin{equation}\label{Eq20}
\Xi_c={{\pi \mu^2} \over {e \hbar}}\  .
\end{equation}
This order of magnitude can easily be understood 
on physical grounds; Schwinger 
discharge is exponentially suppressed unless the work done by 
the electric field on the
virtual pair of (charged) particles 
in separating them by a Compton wavelength
is of the same order
of magnitude (or more) of the particle's mass.
Thus, assuming the existence of elementary particles with mass $\mu$
and charge $e$, a spherical black hole of charge $Q$ 
and radius $r_{+}$ (whose electric field near the horizon 
is $\Xi_{+}= Q / r_{+}^2$) may be considered 
as quasi-static only if it
satisfies the relation $\Xi_{+} \leq \Xi_c$, or equivalently

\begin{equation}\label{Eq21}
|e| \leq {{\pi \mu^2 r_{+}^2} \over {|Q| \hbar}}\  .
\end{equation}
Substituting this into Eq. (\ref{Eq19}) one finds (for $\mu^2 \gg P^2$)

\begin{equation}\label{Eq22}
d\alpha_{min}(s,\mu)={\hbar \over \pi}\  ,
\end{equation}
which is now the fundamental lower bound on the increase in the 
black-hole surface area.
We note that this lower bound is {\it universal} in the sense that it
is {\it independent} of the black-hole parameters $M$ and $Q$.

\section{Summary and Discussion}\label{Sec3}
We have studied the assimilation of a {\it charged} 
particle by a Reissner-Nordstr\"om black hole. The
capture of a particle {\it necessarily} results in an increase in
the black-hole surface area. 
The minimal area increase 
equals to $4 \hbar$. We note that this value is {\it smaller}
than the value given by
Bekenstein for neutral particles. Thus, this process is a better
approximation to a {\it reversible} process in the context of 
black-hole physics.

As was pointed out by Bekenstein
\cite{Beken1} (for neutral particles) the underling physics 
which excludes a completely reversible process is the 
{\it Heisenberg quantum uncertainty principle}. 
However, for {\it charged}
particles it must be supplemented by another physical 
mechanism - {\it a Schwinger-discharge of the black hole}.
Without this mechanism one could have reached the {\it reversible} limit.
It is interesting that the lower bound found here is 
of the same order of magnitude as the one given by
Bekenstein, even-though they emerge from {\it different} 
physical mechanisms.

The {\it universality} of the fundamental lower bound (i.e., 
its independence on the black-hole
parameters $M$ and $Q$) is a further evidence in favor of 
a {\it uniformly} spaced area spectrum for spherical quantum
black-holes (see Ref. \cite{Beken2}). 
Moreover, the universal value $\Delta A_{min}=4 \hbar$ 
is in excellent agreement (to within a factor of
$ln2$) with the area spacing predicted by Mukhanov and 
Bekenstein \cite{MukBek,Beken2} and 
(to within a factor (of order {\it unity}) of
$ln3$) with the area spacing predicted by
Hod \cite{Hod}.

It should be recognized that the precise value of the 
fundamental lower bound Eq. (\ref{Eq22}) can be
challenged. This lower bound follows from 
Eq. (\ref{Eq20}) which can only be interpreted as the critical
electric field to within factors of few.
Nevertheless, the new and interesting 
observation of this paper is the role of {\it Schwinger pair production} in
providing an important limitation on the minimal increase in 
black-hole surface area.

\bigskip
\noindent
{\bf ACKNOWLEDGMENTS}
\bigskip

It is a great pleasure to thank Prof. Tsvi Piran for 
reading the manuscript. 
This research was supported by a grant from the Israel Science Foundation.

\end{document}